\documentclass[floatfix,twocolumn,showpacs,preprintnumbers,amsmath,amssymb,pra,superscriptaddress,longbibliography]{revtex4-1}
\usepackage{color}
\usepackage[usenames,dvipsnames,svgnames,table]{xcolor}
\usepackage[colorlinks=true,linkcolor=blue,urlcolor=blue,citecolor=blue]{hyperref}
\usepackage{mathtools}
\usepackage{graphicx}
\usepackage{dcolumn}
\usepackage{array}
\usepackage{lipsum}
\usepackage{bm}
\usepackage{subfigure}
\usepackage{amssymb}
\usepackage{multirow}
\usepackage{tabularx}
\usepackage{amsmath}
\usepackage{braket}
\graphicspath{{plots/}}
 \usepackage{lipsum}
\usepackage{mathrsfs}
\usepackage{comment}

\newcommand{\beq}{\begin{equation}}
\newcommand{\eeq}{\end{equation}}
\newcommand{\bea}{\begin{eqnarray}}
\newcommand{\eea}{\end{eqnarray}}

\newcommand{\rv}{{\bf r}}
\newcommand{\Rv}{{\bf R}}
\newcommand{\kv}{{\bf k}}
\newcommand{\qv}{{\bf q}}

\newcommand{\ofqwRt}{(\qv,\omega;\Rv,t)}

\newcommand{\retdave}{\text{R}}

\begin{document}

\title{Revealing Non-equilibrium and Relaxation in Warm Dense Matter}

\author{Jan Vorberger}
\email{j.vorberger@hzdr.de}
\affiliation{Helmholtz-Zentrum Dresden-Rossendorf (HZDR), D-01328 Dresden, Germany}

\author{Thomas R.~Preston}
\affiliation{European XFEL, D-22869 Schenefeld, Germany}

\author{Nikita Medvedev}
\affiliation{Institute of Physics, Czech Academy of Sciences, 18221 Prague 8, Czech Republic}
\affiliation{Institute of Plasma Physics, Czech Academy of Sciences, 18200 Prague 8, Czech Republic}

\author{Maximilian~P.~B\"ohme}

\affiliation{Center for Advanced Systems Understanding (CASUS), D-02826 G\"orlitz, Germany}

\affiliation{Helmholtz-Zentrum Dresden-Rossendorf (HZDR), D-01328 Dresden, Germany}

\affiliation{Technische  Universit\"at  Dresden,  D-01062  Dresden,  Germany}

\author{Zhandos A.~Moldabekov}

\affiliation{Center for Advanced Systems Understanding (CASUS), D-02826 G\"orlitz, Germany}
\affiliation{Helmholtz-Zentrum Dresden-Rossendorf (HZDR), D-01328 Dresden, Germany}

\author{Dominik Kraus}
\affiliation{Institut f\"ur Physik, Universit\"at Rostock, D-18057 Rostock, Germany}

\affiliation{Helmholtz-Zentrum Dresden-Rossendorf (HZDR), D-01328 Dresden, Germany}

\author{Tobias Dornheim}
\email{t.dornheim@hzdr.de}
\affiliation{Center for Advanced Systems Understanding (CASUS), D-02826 G\"orlitz, Germany}
\affiliation{Helmholtz-Zentrum Dresden-Rossendorf (HZDR), D-01328 Dresden, Germany}

\begin{abstract}
Experiments creating extreme states of matter almost invariably create non-equilibrium states. These are very interesting in their own right but need to be understood even if the ultimate goal is to probe high-pressure or high-temperature equilibrium properties like the equation of state. Here, we report on the capabilities of the newly developed imaginary time correlation function (ITCF) technique~\cite{Dornheim_T_2022} to detect and quantify non-equilibrium in pump-probe experiments fielding time resolved x-ray scattering diagnostics. We find a high sensitivity of the ITCF even to a small fraction of non-equilibrium electrons in the Wigner distribution. The behavior of the ITCF technique is such that modern lasers and detectors should be able to trace the non-equilibrium relaxation from tens of femto-seconds to several 10s of picoseconds without the need for a model. 
\end{abstract}

\maketitle

\section{Introduction\label{sec:introduction}}

Non-equilibrium comes in many shapes and sizes. Sometimes it is desired, often it needs to be avoided. In any case, its characterization and diagnostics are difficult~\cite{falk_wdm,Macchi_2013}. Arguably even more difficult is the accurate theoretical prediction of a non-equilibrium state and its time evolution~\cite{kremp_book,bonitz_book,Arber_2015,GAMALY201191,new_POP}.

We present here a new method to detect signatures of non-equilibrium in x-ray scattering signals. Inspired by a recent model-free temperature diagnostics in equilibrium~\cite{Dornheim_T_2022}, the method leverages the deviation from the basic detailed balance condition present in equilibrium as measure for non-equilibrium effects. This deviation is calculated based on the imaginary time correlation function (ITCF)  $F(\qv,\tau)$, whose symmetry around its minimum at $\tau=\beta/2$ ($\beta=1/k_BT$) is broken in non-equilibrium. The ITCF can be obtained by a two-sided Laplace transform of the dynamic structure factor $S(\qv,\omega)$. The latter is proportional to the energy resolved x-ray scattering intensity received by a detector positioned at a scattering angle $\theta = 2\arcsin(q/2q_0)$ with $q_0$ denoting the momentum of the incoming photon, $\qv$ and $\omega$ representing the momentum and energy transfer, respectively, during the scattering event~\cite{Dornheim_T_2022,Dornheim_T_follow_up}.

Performing an analysis of the symmetry of the ITCF to detect non-equilibrium relies on nothing but first principles. It does not involve the fitting of the scattering signal with any (approximate) theory nor does it introduce further however well intended approximations. It works for a wide range of probe beams as long as density fluctuations and their correlations are probed and the dispersion of the probe beam in the target is small~\cite{Dornheim_T_follow_up}. This is the case for hard and soft x-rays, electron beams, and even neutrons. As we will show below, the method can work using a single detector at a single scattering angle but cross checks with several detectors at a few different angles provide additional information and certainty in the equilibrium/non-equilibrium assessment. The shapes of well-characterized source and probe functions can be taken into account without difficulty.

Leveraging the capabilities of modern optical and x-ray laser systems allows to not only perform one measurement but indeed track and characterize the entire time evolution of the non-equilibrium in the driven state and later during the relaxation phase~\cite{Tschentscher_2017,LCLS_2016,SACLA_2011}. This allows first principles access to relaxation times and energy transfer rates between subsystems that have hitherto been restricted by their strong reliance on two-temperature models~\cite{Waldecker_2015}, amongst others. 

The intended use for this method is for experiments involving the creation and characterization of warm dense matter and other extreme states of matter~\cite{Knudson_Science_2015,Fletcher2015,Kraus2016,Kraus2017,Celliers_Science_2018,Lazicki2021,Gaus_2021}. Such states are transient in the laboratory but are commonplace in astrophysical settings like planetary interiors~\cite{Liu2019,Brygoo2021,Militzer_2022}, dwarf star interiors~\cite{Kritcher_Nature_2020,becker}, neutron star crusts~\cite{Chamel2008}, and the like. It is of paramount importance to know the relaxation times of a variety of elements and materials after shock absorption or laser irradiation to study equilibrium properties of said materials to understand their behaviour at extreme pressures for the earlier mentioned planetary settings~\cite{White_2020}. In particular, experiments using powerful x-ray laser beams to isochorically heat and at the same time probe the target should analyse their scattering signal using the method presented here~\cite{Sperling_PRL_2015,kraus_xrts}. 

The investigation of creation and relaxation of hot electrons in materials at ambient pressure conditions in solid state physics~\cite{Zahn_ni_2021,Zahn_co_ni_2022} drives many technological applications in chip manufacturing~\cite{MARKS_2022}, medicine~\cite{VV_2003}, surface modification~\cite{FW_2006}, or generally laser materials processing~\cite{bauerle_book,BS_2013}. From a basic research point of view, we are interested in the general coupling of the ionic and electronic degrees of freedom in (non-)equilibrium~\cite{transfer1,transfer2,simoni_2020,Medvedev_2020,zhang_2022, CiricostaModel} and a great many experiments have been performed to test the theoretical predictions for non-equilibrium warm dense matter states~\cite{CNXF_1992,WVBC_2012,ZSHB_2014,HBCD_2015,Mo_gold_2018,Fletcher_2022,Vinko2012,Ciricosta2012,Ciricosta2016,Preston2017}. We aim to understand the microscopic dynamics governing the energy exchange between, for example, $\alpha$-particles and electrons and between electrons and protons/deuterons/tritons in the fusion pellet of inertial confinement fusion~\cite{Edie_2013,Zylstra2022}. Such questions are naturally related to the study of response functions as they are of further importance for the stopping power~\cite{Malko2022}, transport quantities like conductivities~\cite{Ramakrishna_2022}, or nonlinear effects like higher harmonics generation~\cite{Dornheim_PRL_2020,Dornheim_JCP_ITCF_2021,Dornheim_JPSJ_2021,Dornheim_PRR_2021,Moldabekov2022}. 

To demonstrate the capabilities of our method, we first compute theoretical dynamic structure factors and ITCFs for very specific well defined non-equilibrium Wigner distributions. This is done using the technique of real-time Green's functions~\cite{kremp_book,Chapman_2011,Vorberger_PRE_2018}. This gives insight into some general behavior of the ITCF for typical non-equilibrium situations. We then demonstrate how the time evolution of the non-equilibrium state may be tracked in an experiment using ITCFs. For this, we use solutions for the time evolution of the Wigner function from a Boltzmann-Monte Carlo code~\cite{Medvedev2022}.

Since the set of possible non-equilibrium situations in pump-probe scenarios is almost infinite, we restrict the discussions in this paper to systems in non-equilibrium that are homogeneous and isotropic. This is a substantial restriction but leaves the electron Wigner distribution function free to take any form. A generalization to in-homogeneous or anisotropic systems~\cite{Chapman_2014,Kozlowski_2016} is possible and will be covered in a future publication.

\section{Theory\label{sec:theory}}

\subsection{Theory for the non-equilibrium structure factor\label{subsec:sqwtheory}}

The structure factor $S$ is defined by the correlation function of density fluctuations $L^>$ in non-equilibrium according to~\cite{kremp_book}
\bea\label{sqwrt}
S_{ab}\ofqwRt&=&\frac{i}{2\pi}L_{ab}^{>}\ofqwRt\\
&=&\frac{1}{2\pi \sqrt{n_an_b}}\int d\rv ds\;
e^{i(\omega s-\qv\cdot\rv)} i\hbar L_{ab}^{>}(12)\,,\nonumber
\eea
where $1=\{\rv_1,t_1,\sigma_1\}$ denotes a complete set of observables and the species are denoted by the labels $\{a,b\}$. 
Here, a Fourier transform is performed with respect to the relative coordinates in space and time $\rv=\rv_1-\rv_2$ and $s=t_1-t_2$, respectively. The center of mass variables $\Rv=(\rv_1+\rv_2)/2$ and $t=(t_1+t_2)/2$ remain untransformed. Thus, the microscopic fluctuations are considered in momentum-frequency space whereas the macroscopic variations in density etc.~are treated in real space. For instance, the system can still be homogeneous macroscopically but the Wigner distribution may indicate severe deviations from Fermi- or Boltzmann distributions. In this case, the dependencies on $\Rv$ and $t$ are dropped.

The correlation function of density fluctuations $L_{ab}^>$ is given by~\cite{kremp_book}
\beq
i\hbar L_{ab}^>(12)=\langle\delta\rho_a(1) \delta \rho_b(2)\rangle\,.
\eeq
The average is performed as trace over the fluctuation operators and the non-equilibrium density operator $\hat{\rho}$, $\langle\ldots\rangle=\mbox{Tr}\{\hat{\rho}\ldots\}$. The density fluctuations are given in coordinate space as
\beq
\delta\rho(1)=\psi^{\dagger}(1)\psi(1)-\langle\psi^{\dagger}(1)\psi(1)\rangle\,,
\eeq
where we used the creation and annihilation operators $\psi^{\dagger}$ \& $\psi$ in second quantization.

We still define an intermediate scattering function in imaginary time $\tau$ (ITCF) with a relation similar to Eq.~(\ref{sqwrt})
\beq\label{itcf}
F_{ab}(\qv,\tau;\Rv,t)=\int \limits_{-\infty}^{\infty} d\omega\;
e^{-\omega\tau} S_{ab}\ofqwRt\,.
\eeq
This intermediate scattering function has, in equilibrium, all the known symmetries, in particular detailed balance, but will show significant deviations from them in a non-equilibrium case which can be used to decide the degree of non-equilibrium. We introduce
\beq
F_{ab}(\qv,\tau)/F_{ab}(\qv,\beta'-\tau)
\label{deviation}
\eeq
as such a measure for non-equilibrium. Here $\beta'$ is defined by $F(\tau=0)=F(\tau=\beta')$. The quantity of Eq.~(\ref{deviation}) will approach unity the closer the system gets to equilibrium and $\beta'$ will be the inverse temperature then.

Of particular interest is of course the total electron-electron structure factor $S_{ee}$ and its ITCF as it can be obtained in a variety of scattering experiment, e.g. via x-ray Thomson scattering~\cite{siegfried_review}. However, noise in the experimental signal, a finite energy bandwidth of the probe beam, and in the detector function, and a finite range of energies available in the experiment ($[-x,x]$) need to be accounted for
\beq\label{itcf_exp}
F_{ee}^{exp}(\qv,\tau;\Rv,t)=\int \limits_{-x}^{x} d\omega\;
e^{-\omega\tau} I_{ee}\ofqwRt\,,
\eeq
where 
\beq
I_{ee}\ofqwRt=R(\omega)\otimes S_{ee}\ofqwRt+\Delta I\ofqwRt
\eeq
is the convolution of the detector and source function [summarized as $R(\omega)$] with the dynamic structure factor and Gaussian noise $\Delta I$ added~\cite{sheffield2010plasma,Dornheim_T_2022,Dornheim_T_follow_up}.

Explicit expressions for the dynamic structure in non-equilibrium can be obtained from the equation of motion for the correlation function of density fluctuations~\cite{Vorberger_PRE_2018}. Sometimes these relations are called generalized fluctuation-dissipation theorems~\cite{kremp_book}.
Thus, for a realistic system comprised of electrons and ions, the non-equilibrium total electron-electron structure factor is in direct polarization approximation (DPA, neglecting cross-species polarization functions) given by~\cite{Vorberger_PRE_2018} 
\begin{align}
\label{Lee_no_pi_ei}
  S_{ee}\ofqwRt=\frac{1}{2\pi n_e}L_{ee}^{>}
  = &\,
  \frac{{\cal L}_{ee}^{>} + |{\cal L}_{ee}^{\retdave}\!|^2V_{ei}^2{\cal L}_{ii}^{>}}
  {|1-V_{ie}{\cal L}_{ee}^{\retdave} V_{ei}{\cal L}_{ii}^{\retdave}|^{2}}\,,
\end{align}
Here and below, we omit all the arguments $\ofqwRt$ of every quantity on the right hand sides. The pure species contribution appearing above are defined by
\begin{align}
{\cal L}_{aa}^>\ofqwRt=&\frac{\Pi_{aa}^> }
{|1 - \Pi_{aa}^{\retdave}V_{aa} |^2}\,,
\label{pol_funcs_aa_>/<}
\end{align}
for the greater functions and the retarded functions are given by
\begin{align}
{\cal L}_{aa}^{\retdave}\ofqwRt=&\frac{\Pi_{aa}^{\retdave}}{1 - \Pi_{aa}^{\retdave}V_{aa}}\,.
\label{pol_funcs_aa_R/A}
\end{align}
The Coulomb potential is denoted by $V$.
The polarization functions $\Pi$ are needed as input quantities and they can be computed in random phase approximation (RPA) or using non-equilibrium local field corrections~\cite{Vorberger_PRE_2018}. The RPA is sufficient to demonstrate the capabilities of the intermediate scattering function in imaginary time. The retarded polarization function $\Pi^R$ may be obtained from the correlation functions and a  Kramers-Kronig relation
\bea
\mbox{Im}\Pi^R\ofqwRt&=&\frac{i}{2}\left(\Pi^<\ofqwRt\right.
\nonumber\\
&&\qquad\left.-\Pi^>\ofqwRt\right)\,,\\
\mbox{Re}\Pi^R\ofqwRt&=&\int\frac{d\omega'}{2\pi}
\frac{\mbox{Im}\Pi^R(\qv,\omega';R,t)}{\omega-\omega'}\,.
\eea
The correlation functions themselves can easily be obtained in RPA from
\bea
\Pi^{\lessgtr}\ofqwRt&=&2\pi i\int\frac{d\kv}{(2\pi\hbar)^3}
f^{\lessgtr}(\kv+\qv,t)f^{\gtrless}(k,t)\nonumber\\
&&\times\delta\left(\hbar\omega-E(\qv+\kv)+E(k)\right)\,.
\eea
The energy dispersion can be taken from the ideal gas $E(k)=(\hbar k)^2/2m$ and the Wigner distributions are $f^>=-[1-f(k,t)]$ and $f^<=f(k,t)$.

Within a two-temperature model, Eq.~(\ref{Lee_no_pi_ei}) can be used to determine not only the electron temperature but also the ion temperature as the DPA is valid for a two temperature system. The ITCF defined in Eq.~(\ref{itcf}) will indicate the correct electron temperature due to the large mass ratio between electrons and ions. Knowledge of the electron temperature allows to solve Eq.~(\ref{Lee_no_pi_ei}) for the pure ion structure ${\cal L}_ {ii}$ whose temperature can then be determined via the ion ITCF.

\subsection{Theory for the Wigner function\label{subsec:wignertheory}}

The non-equilibrium dynamic structure factor $S\ofqwRt$ depends on the knowledge of the Wigner function, which itself is non-trivial to obtain via a kinetic equation~\cite{kremp_book,bonitz_book}. For demonstration purposes, we can model the non-equilibrium (electron) Wigner function by a bump-on-hot-tail distribution that has been found to occur in various laser pump-probe setups~\cite{Faustlin2010,Chapman_2011}
\bea
f(k,t)&=&A_c \left\{\exp\left[\beta_c\left(\frac{\hbar^2k^2}{2m}-\mu_c\right)\right]+1\right\}^{-1}\nonumber\\
&+&A_h\left\{\exp\left[\beta_h\left(\frac{\hbar^2k^2}{2m}-\mu_h\right)\right]+1\right\}^{-1}\nonumber\\
&+&A_b \frac{n\Lambda^3}{2}\exp\left(-\beta_b\frac{(\hbar k-p_b)^2}{2m}\right)\,.
\eea

Here, the indices $\{c;h;b\}$ stand for {\em cold}, {\em hot}, and {\em bump}, respectively. The {\em bump} is due to direct laser irradiation and is located at momentum $p_b$. The thermal wavelength is $\Lambda^2=2\pi\hbar/mk_BT_b$. Each part of the Wigner function has its own inverse 'temperature' $\beta_i$, and chemical potential $\mu_i$. The latter two and also the fractions $A_i$ can change over time. The total density of electrons is given by the integral over all momenta, $\hbar k$.

This shape of the transient distribution function in X-ray-irradiated materials was confirmed by various simulations of the non-equilibrium electronic system~\cite{Faustlin2010,Medvedev2011a,Hau-Riege2013}.
We use an event-by-event Monte Carlo (MC) simulation to trace non-equilibrium evolution of the electron distribution function, augmented with Pauli blocking factors in the scattering probability calculations (which is essentially identical to a numerical solution of the semi-classical Boltzmann equation)~\cite{Medvedev2011a}. The method uses an individual particle simulation to model the response of the electronic system of aluminium to vacuum ultra-violet or soft X-ray irradiation. The photoabsorption cross sections are extracted from Ref.~\cite{Henke1993}. In case photoabsorption excites a core-shell, its Auger decay then promotes another electron to the conduction band. The auger decay times are taken from Ref.~\cite{Keski-Rahkonen1974}. All electrons in the conduction band, excited and existing in aluminium prior to irradiation, are then scattering with the probabilities depending on the cross section of scattering multiplied with factors containing the transient distribution function, in accordance with the Boltzmann collision integral~\cite{Medvedev2011a}. The cross sections of free-free electron scattering is approximated within the linear response theory with the Mermin-like complex dielectric function for aluminium~\cite{Medvedev2022}. Tracing each scattering event of each electron within the selected simulation box allows to model the non-equilibrium evolution of the electron distribution function in time. Despite the fact that electrons in the Monte Carlo scheme are classical particles, inclusion of the Pauli blocking factors in the scattering probabilities allows us to obtain a quantum distribution function (i.e. Fermi-Dirac instead of Maxwell-Boltzmann in the case of equilibrium), see Ref.~\cite{Medvedev2011a} for details. We use 10$^4$ MC iterations with approximately 3000 particles in each to obtain reliable statistics.

\section{Analysis of scattering spectra}

We analyze the effects of different types of idealized non-equilibrium situations. We restrict ourselves to isotropic and homogeneous conditions in the system. Macroscopic gradients in the density can be taken into account via a gradient expansion of the microscopic variables~\cite{Chapman_2014,Kozlowski_2016}.

\subsection{Study of idealized non-equilibrium situations}

\begin{figure*}
\centering
\includegraphics[width=\textwidth]{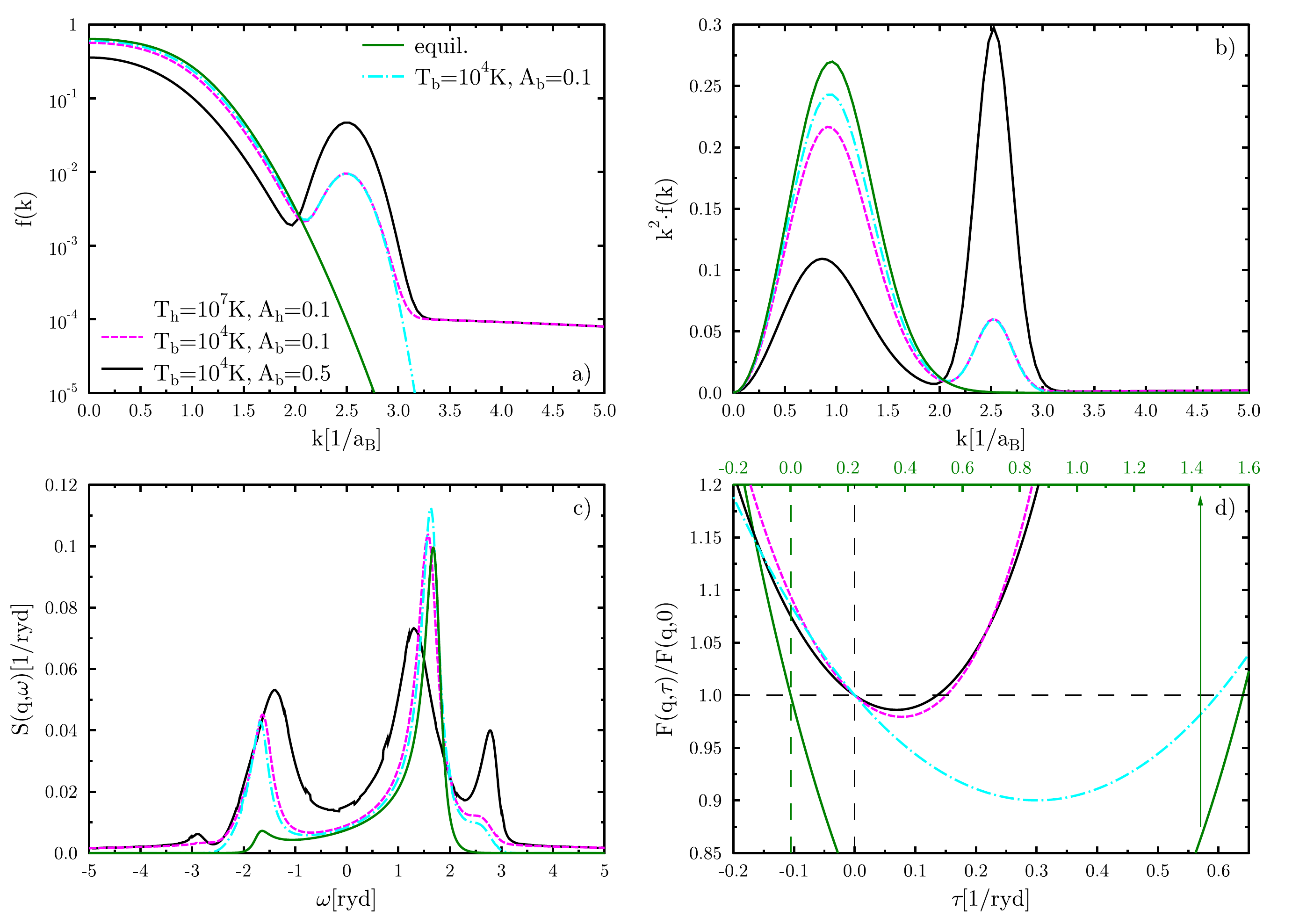}
\caption{\label{fig:itcf_theo_1} Non-equilibrium structure factor of an electron gas in a VUV-pump scenario. The Wigner distribution function is shown in panel a); panel b) shows the Wigner function multiplied with the square of the momentum to better visualize the fractions of electrons at certain momenta. The dynamic structure factor is shown in panel c). The intermediate scattering function is shown in panel d). The electronic density is $n=1.8\times 10^{23}$~cm$^{-3}$ ($r_s=2.07$). The equilibrium temperature (cold bulk) is $T_c=10^5$~K. The hot tail has a temperature of $T_h=10^7$~K. The wavenumber is $q=0.5/$~a$_B$. In panel d), the x-axis for the equilibrium case is on top, the non-equilibrium x-axis is at the bottom. 
}
\end{figure*}
We first study non-equilibrium effects in an uniform electron gas, perhaps the most important model system in plasma physics, warm dense matter physics and solid state physics~\cite{dornheim_prl,dornheim_POP,review,dornheim_dynamic,dornheim_ML,Dornheim_PRL_2020,Dornheim_PRE_2021,Dornheim_PRB_2021,Dornheim_PRR_2021,Dornheim_Nature_2022}.

A VUV laser of moderate intensity produces a bump in the electron Wigner distribution function in the vicinity of the contribution of the cold bulk electrons. Such an example is presented in Fig.~\ref{fig:itcf_theo_1}. We show different (substantial) fractions of the electrons pumped from the cold bulk into the bump.  The deviations from the equilibrium structure factor (solid green line) are easily noticeable in panel c) of Fig.~\ref{fig:itcf_theo_1}.  As the chosen wavenumber lies within the collective regime, all structure factors feature plasmon-like peaks. However, the position of the non-equilibrium plasmon peaks is shifted to lower energies as compared to the equilibrium case. This is caused by the lowering of the bulk density when electrons are laser heated into a hot tail or bump. Hot electrons contribute less to screening and collective effects. In addition, the ratio of the plasmon peaks is changed considerably as the detailed balance symmetry is broken in non-equilibrium. The bump causes in addition further peaks of the dynamic structure factor at higher energies. These might be identified as beam acoustic modes. 

Fig.~\ref{fig:itcf_theo_1} panel d) displays the ITCFs for all the cases. The green ITCF for the equilibrium case is symmetric around $\beta/2$ and shows the correct temperature. Analyzing the ITCFs of the three non-equilibrium cases gives spurious high temperatures of $T^*=2.5\times 10^5$~K and above. As the signatures of non-equilibrium are very obvious when looking at the dynamic structure factors, the ITCFs seem very equilibrium like in the intervals $[0,\beta^*]$. Mirroring the ITCFs around the $\beta^*/2$-axis would only reveal deviations from the equilibrium symmetry far away from $\beta^*/2$. This means that the collective regime is not well suited to distinguish equilibrium from non-equilibrium via the ITCF. As long as most of the weight of the dynamic structure factor is contained in plasmon-like peaks, the ITCF will always be symmetric showing some spurious temperature stemming from the ratios of the non-equilibrium plasmon peaks. This can easily be demonstrated when a simple plasmon model for the dynamic structure factor is Laplace-transformed into the respective ITCF.

\begin{figure}
\centering
\includegraphics[width=0.48\textwidth]{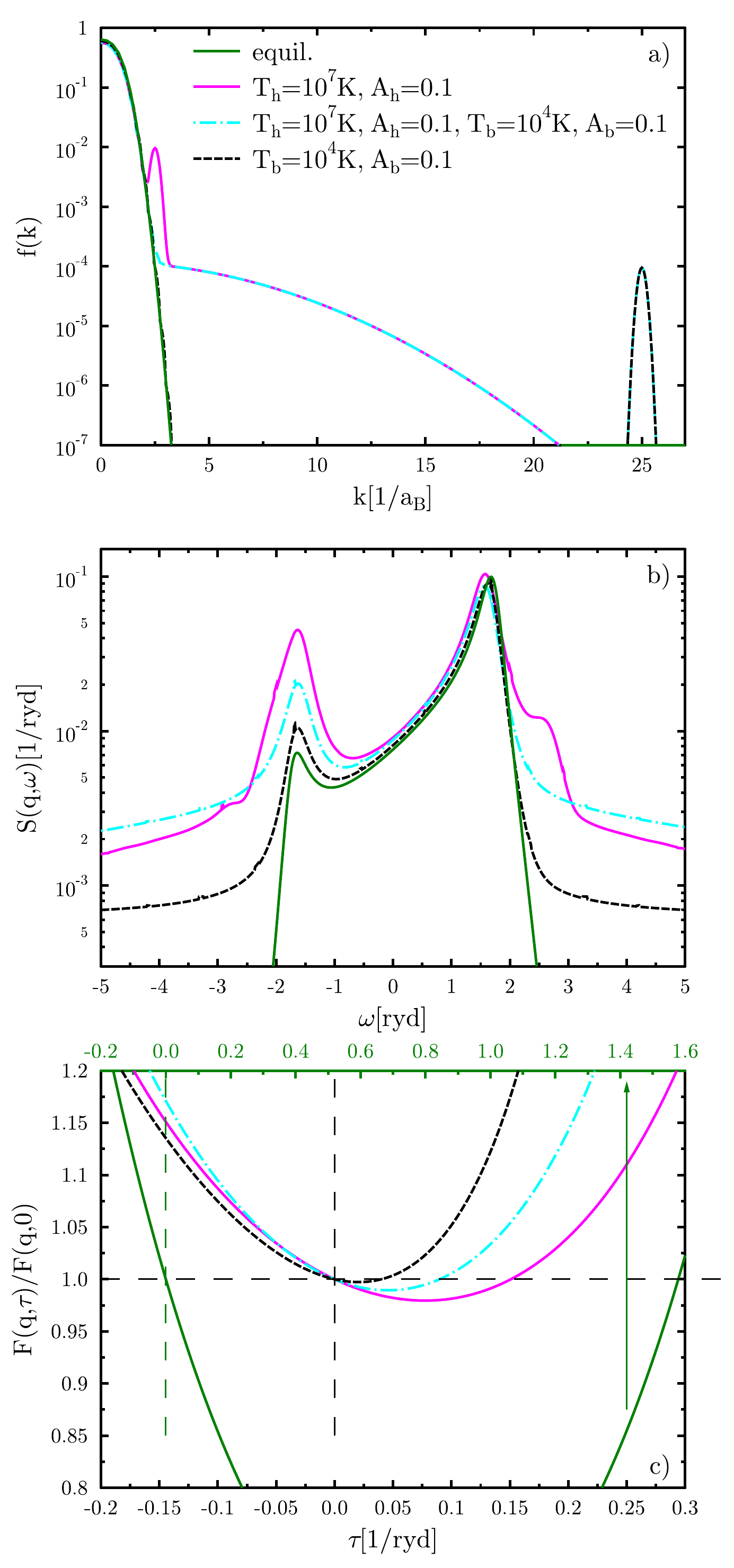}
\caption{\label{fig:itcf_theo_2} Non-equilibrium structure factor of an electron gas comparing a VUV and a XFEL-pump scenario. The Wigner distribution function is shown in panel a). The dynamic structure factor is shown in panel b). The intermediate scattering function is shown in panel c). The electronic density is $n=1.8\times 10^{23}$~cm$^{-3}$. The equilibrium temperature (cold bulk) is $T_c=10^5$~K. The hot tail has a temperature of $T_h=10^7$~K. The wavenumber is $q=0.5/$~a$_B$. In panel c), the x-axis for the equilibrium case is on top, the non-equilibrium x-axis is at the bottom.}
\end{figure} 
Next we want to analyse the effect of different bump energies, i.e., the effect of pumping the system with harder and harder laser radiation like x-rays instead of VUV. An example is presented in Fig.~\ref{fig:itcf_theo_2}. The case with the solid magenta curve is the same as in Fig.~\ref{fig:itcf_theo_1} featuring a hot tail and a bump at VUV energies. Retaining only the hot tail, but removing the bump (as might be the case after the pump laser has switched off, solid cyan line), makes the additional mode around $\omega=3$~ryd vanish (1~ryd$ = 13.6$\,eV). Only visible due to the log scale in Fig.~\ref{fig:itcf_theo_2}, the other main difference between equilibrium and the considered non-equilibrium cases, apart from the differences in collective mode peaks, is the behaviour of the high energy tails. In the (solid green) equilibrium case, the dynamic structure factor shows exponential decay around $\omega=\pm 2$~ryd. In the non-equilibrium cases, the high energy electrons in the Wigner distribution cause a much slower decay and high energy plateaus of the dynamic structure factor (the exponential decay sets in around $\omega=\pm 25$~ryd in the x-ray case (black line). Further, the higher energy bump in the Wigner distribution at x-ray energies (see blacked dashed line in panel a) of Fig.~\ref{fig:itcf_theo_2}, $k=25/$a$_B$ is equivalent to $8.5$~keV pump laser energy), even though it is responsible for the large tails, has a lot less influence on the plasmon peaks than the VUV bump or the hot tail. These electrons are energetically too far removed to significantly influence the collective dynamics. We have already seen that under VUV excitation, the asymmetry in the ITCF is small, such that the collective regime cannot be used to detect non-equilibrium from the ITCF. However, for the x-ray laser excitation case (black dashed line in panel c) of Fig.~\ref{fig:itcf_theo_2}), such asymmetry is visible and a single detector at a single scattering angle should be sufficient.

\begin{figure}
\centering
\includegraphics[width=0.48\textwidth]{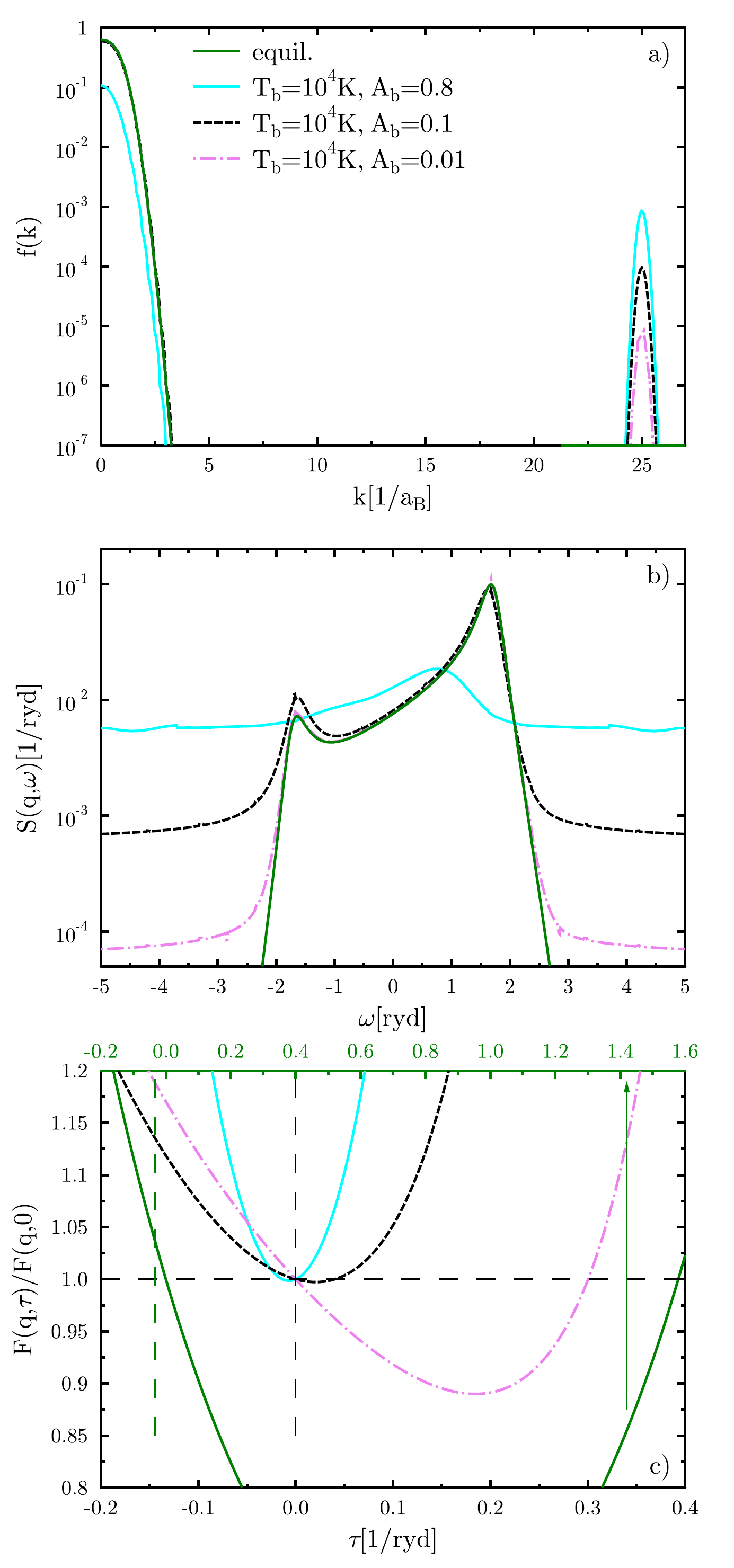}
\caption{\label{fig:itcf_theo_3} Non-equilibrium structure factor of an electron gas in a XFEL-pump scenario. The Wigner distribution function is shown in panel a). The dynamic structure factor is shown in panel b). The intermediate scattering function is shown in panel c). Only the electrons are considered. The electronic density is $n=1.8\times 10^{23}$~cm$^{-3}$. The equilibrium temperature (cold bulk) is $T_c=10^5$~K.  The wavenumber is $q=0.5/$~a$_B$. In panel c), the x-axis for the equilibrium case is on top, the non-equilibrium x-axis is at the bottom.}
\end{figure} 
The question then naturally arises, how big the fraction of electrons need to be in the x-ray pump laser scenario for the ITCF symmetry to notice the non-equilibrium. Such a case study is presented in Fig.~\ref{fig:itcf_theo_3}. Apart from the usual equilibrium case (solid green line), we show cases with $1\%$, $10\%$, and $80\%$ of the electrons being contained in the bump at $k=25/a_B$, respectively. Provided the high energy tails of the dynamic structure factor beyond $\omega=\pm 3$~ryd can be resolved and detected in experiment, even a $1\%$ population of electrons at $8.5$~keV can be detected in the ITCF in the collective regime, see panel c). This is even more remarkable, as is clearly visible in panel b) of Fig.~\ref{fig:itcf_theo_3} that the dynamic structure is very similar to the equilibrium case for energies in the range $\omega=\pm 2$~ryd (dash-dotted violet curve). Naturally, if substantial amounts of electrons are pumped out of the bulk distribution, the asymmetry in the ITCF becomes more and more obvious (black dashed and solid cyan curves).

\begin{figure}
\centering
\includegraphics[width=0.48\textwidth]{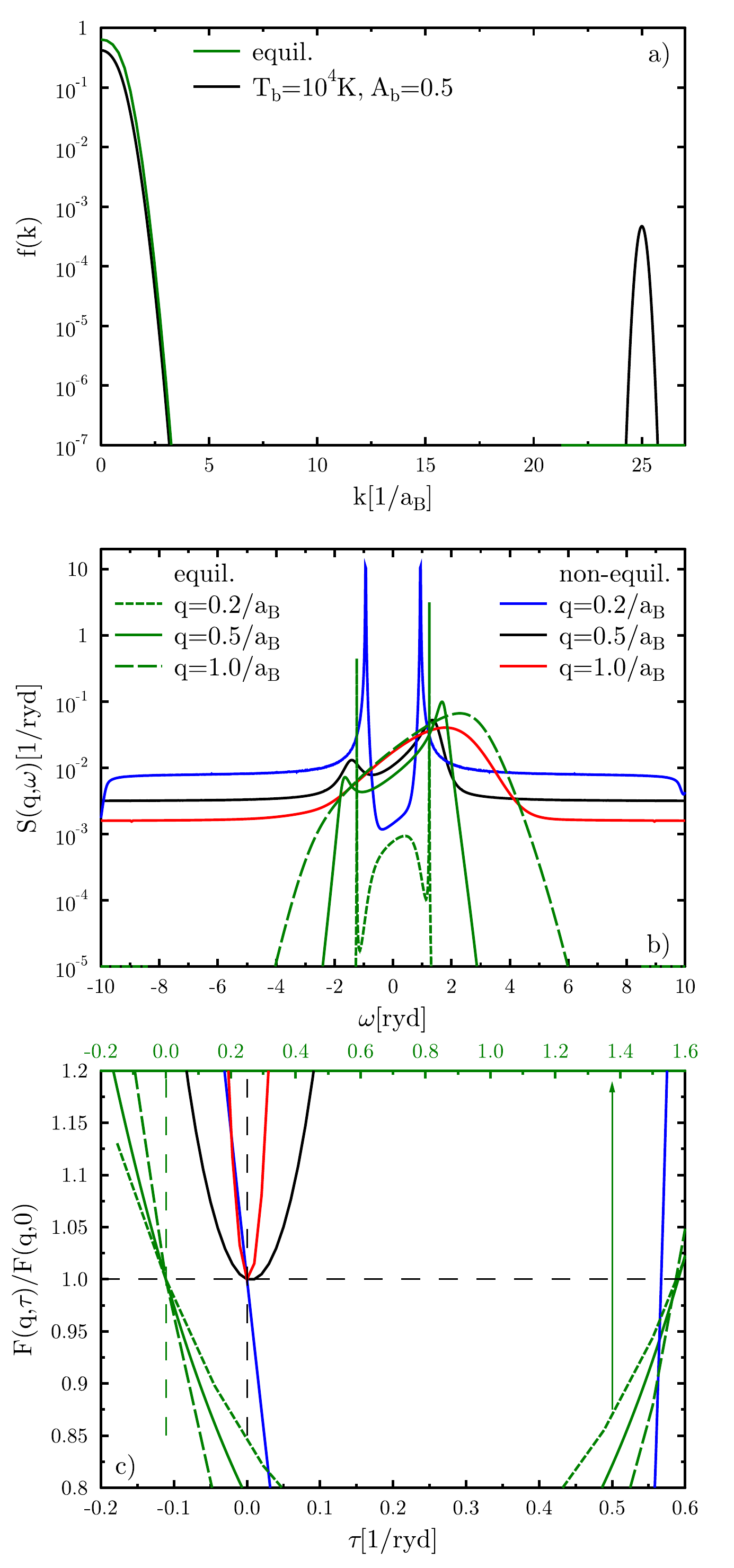}
\caption{\label{fig:itcf_theo_4} Non-equilibrium structure factor of an electron gas in a XFEL-pump scenario for different wavenumbers. The Wigner distribution function is shown in panel a). The dynamic structure factor is shown in panel b). The intermediate scattering function is shown in panel c). The electronic density is $n=1.8\times 10^{23}$~cm$^{-3}$. The equilibrium temperature (cold bulk) is $T_c=10^5$~K. In panel c), the x-axis for the equilibrium case is on top, the non-equilibrium x-axis is at the bottom.}
\end{figure} 
In case the experimental resolution is insufficient or the asymmetry in the ITCF is too small to be conclusive, the logical next step would be to check the dynamic structure and therefore the ITCF at different scattering angles, i.e., different wavenumbers. This has been done in Fig.~\ref{fig:itcf_theo_4}. Only a single non-equilibrium distribution as shown in panel a) is evaluated at three different wavenumbers. These cover the collective, transitional and single particle regimes. In the case of equilibrium, see the green curves, all ITCFs converge at the same temperature. Plotting the ITCFs for the x-ray bump scenario (blue, black, and red curves) gives (apart from any asymmetry in the ITCF) three totally different values for their spurious temperatures. This constitutes an unambiguous experimental signature of non-equilibrium. It is again interesting to observe how the high energy tails behave with increasing wavenumber. Even though the tail for the largest wavenumber considered here is the lowest of the three cases in the energy range shown in Fig.~\ref{fig:itcf_theo_4}, it extends farthest out. In the collective (blue) regime, the high energy tail ends at around $\omega= \pm 10$~ryd, in the transitional regime (black) it extends to $\omega= \pm 25$~ryd, but in the single particle regime (red), it stretches to $\omega= \pm 50$~ryd at the same level of intensity as shown here. In this case, it would mean that a proper characterization of the target would require the detector to be highly sensitive in the energy range from $7800$~eV to $9200$~eV with an x-ray XFEL beam at $8500$~eV. This energy range has been demonstrated at the HED instrument of the European XFEL~\cite{Preston2020}.

\begin{figure}
\centering
\includegraphics[width=0.48\textwidth]{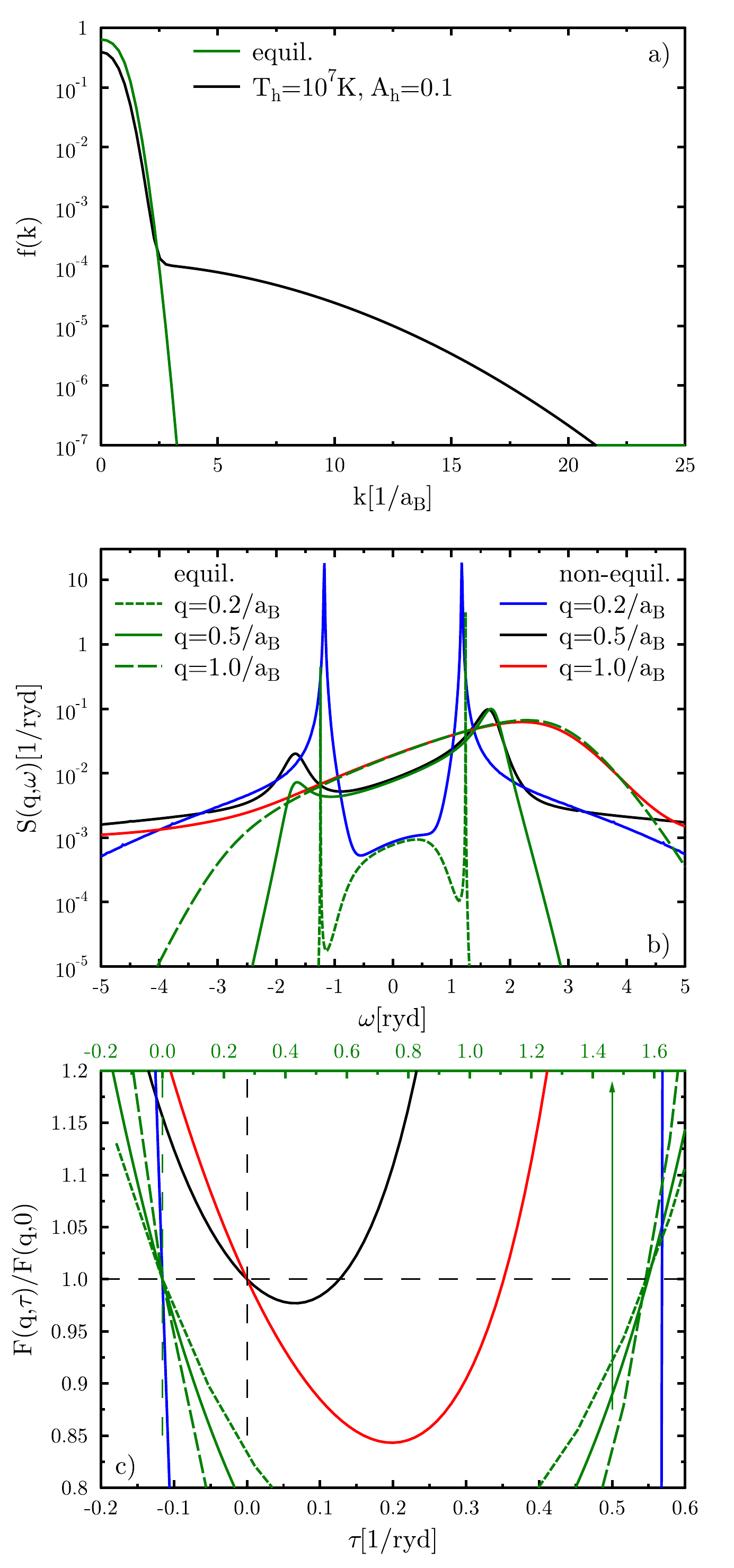}
\caption{\label{fig:itcf_theo_5} Non-equilibrium structure factor of an electron gas in a two-temperature scenario. The Wigner distribution function is shown in panel a). The dynamic structure factor is shown for different wavenumbers in panel b). The intermediate scattering function is shown in panel c). The electronic density is $n=1.8\times 10^{23}$~cm$^{-3}$. The equilibrium temperature (cold bulk) is $T_c=10^5$~K. The temperature of the tail is $T_h=10^7$~K. In panel c), the x-axis for the equilibrium case and the blue non-eq. case is on top, the non-equilibrium x-axis is at the bottom. 
}
\end{figure} 
After the pump laser is switched off, the electrons relax back towards equilibrium. One such intermediate state could be a scenario in which there are two populations of electrons with different temperatures as presented in Fig.~\ref{fig:itcf_theo_5}. Again, we consider three different wavenumbers so as to not only depend on the analysis of the symmetry properties of the ITCF. Again, we can observe large high energy tails due to the high temperature tail in the Wigner distribution. As before, even if asymmetries in the ITCFs cannot be resolved, the differences in the detected temperatures for different wavenumbers are such that only a single conclusion is possible.

\begin{figure*}
\centering
\includegraphics[width=\textwidth]{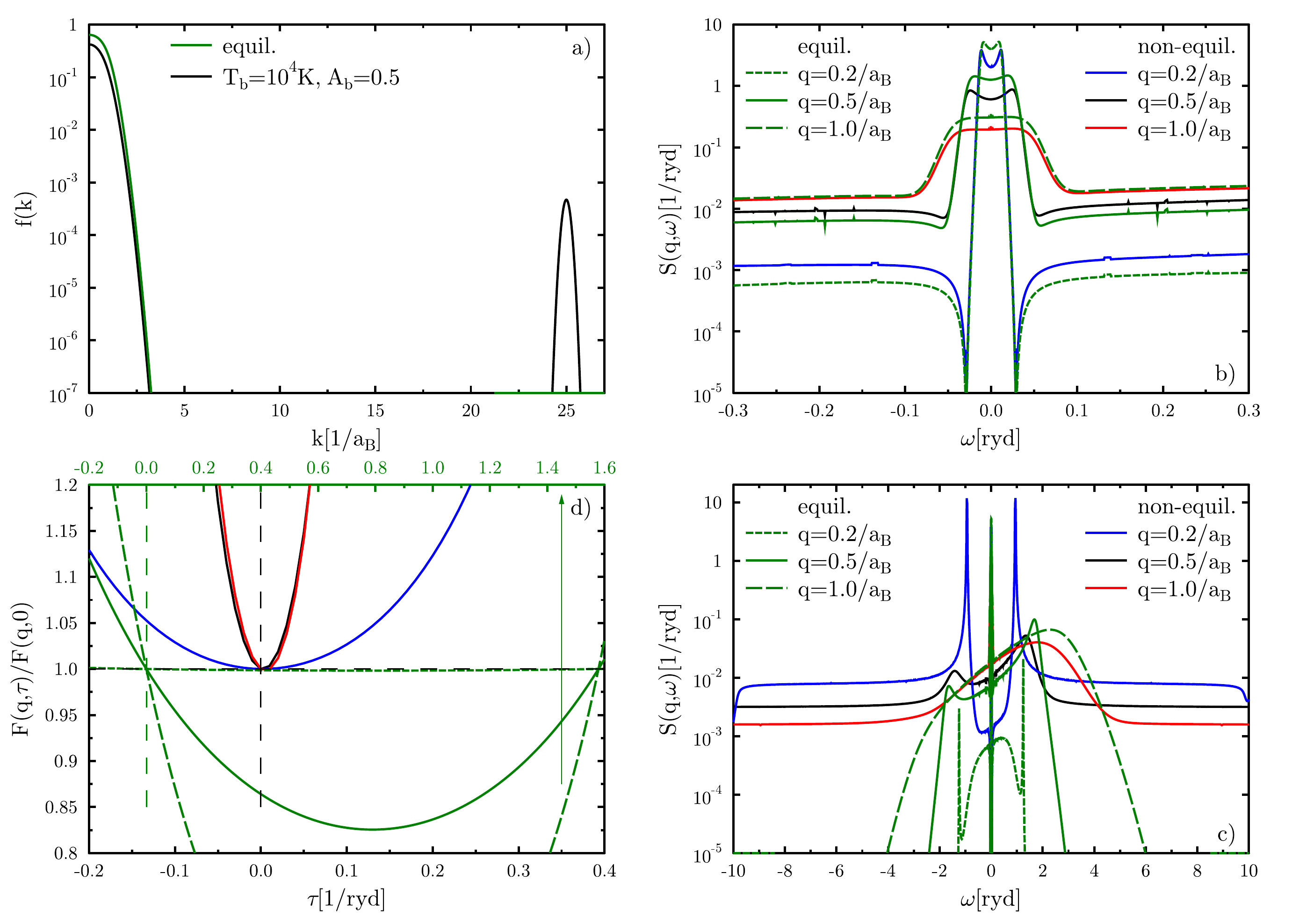}
\caption{\label{fig:itcf_theo_6} Non-equilibrium electron structure factor in fully ionized hydrogen. The total electronic density is $n=1.8\times 10^{23}$~cm$^{-3}$. The temperature of the cold bulk electrons is $T_c=10^5$~K. The protons have a temperature of $T=10^5$~K. The Wigner distribution function is shown in panel a).  The dynamic structure factor for the phonon and plasmon energy ranges is shown in panels b) \& c), respectively. The intermediate scattering function is shown in panel d). In panel d), the x-axis for the equilibrium case is on top, the non-equilibrium x-axis is at the bottom.}
\end{figure*} 
After analysing the behaviour of the pure electron gas in several different non-equilibrium situations, we now focus on a more realistic but relatively simple system, namely fully ionized hydrogen. As a real material, it will not only feature the signal of the electrons in the dynamic structure factor, such as plasmon peaks, but the ion acoustic modes. The latter will be modified in appearance and occurrence from the equilibrium situation. An example is given in Fig.~\ref{fig:itcf_theo_6}. Panel a) contains the familiar situation of an x-ray pumped electron distribution. Panel b) focuses on the ion modes. It is clear that the reduced number of bulk electrons change the screening of the ion modes. We observe that the modes are more pronounced in the non-equilibrium collective regime when comparing them to the equilibrium results. On the other hand, the electron modes and peaks are hardly altered by the presence of the equilibrium protons, see panel c) and compare to Fig.~\ref{fig:itcf_theo_4}. Nevertheless, the analysis of the ITCFs for the three different wavenumbers shows strong deviations in behavior and symmetry. No minimum in the ITCFs can be seen with the naked eye making the distinction as non-equilibrium simple, even without the use of three different wavenumbers.

\begin{figure*}
\centering
\includegraphics[width=\textwidth]{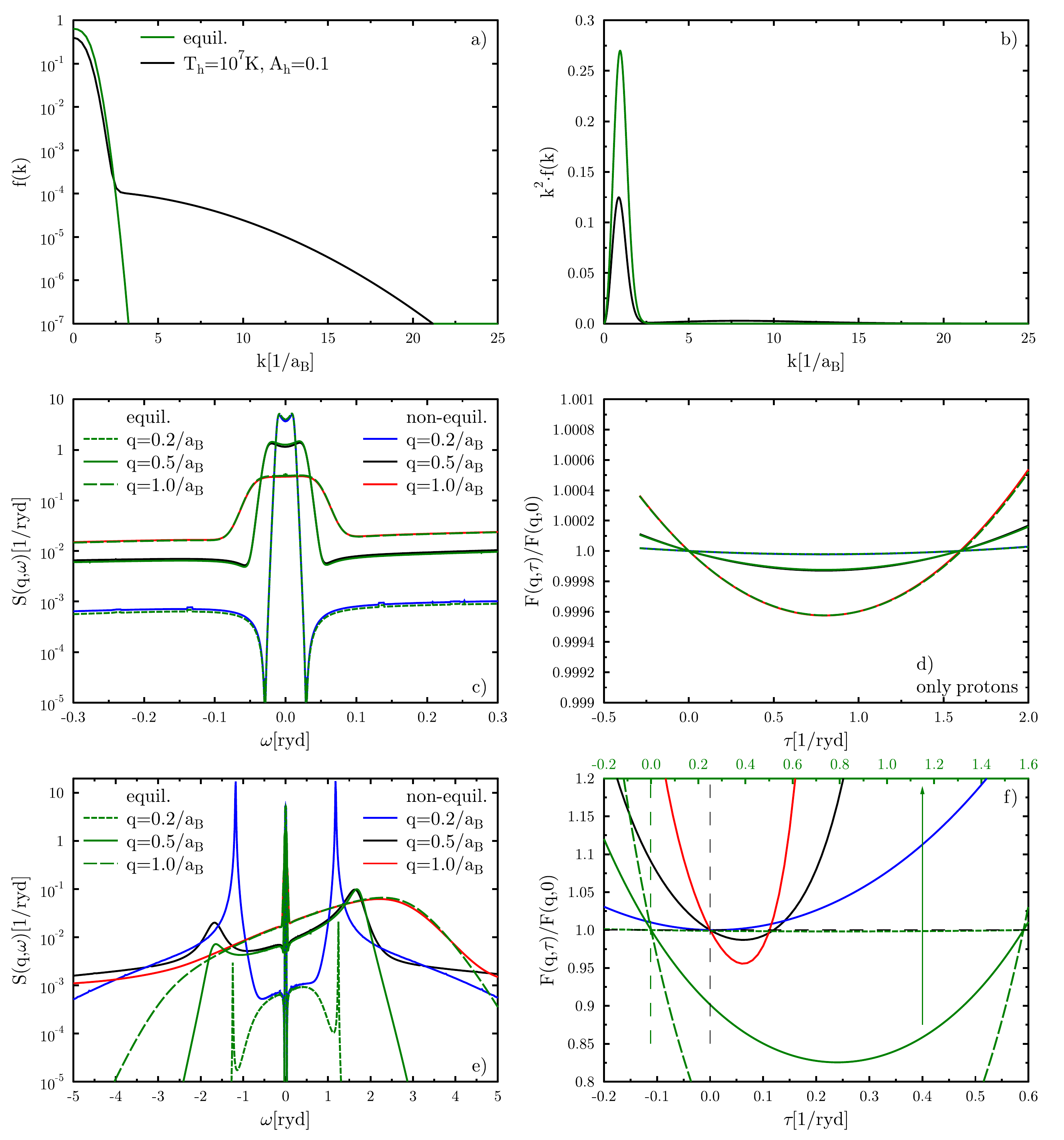}
\caption{\label{fig:itcf_theo_7} Non-equilibrium electron structure factor in a two-temperature scenario in fully ionized hydrogen. The electronic density is $n=1.8\times 10^{23}$~cm$^{-3}$. The temperature of the cold bulk electrons is $T_c=10^5$~K. The temperature of the electrons in the tail is $T_h=10^7$~K. The protons have a temperature of $T=10^5$~K. The Wigner distribution function is shown in panels a) \& b).  The dynamic structure factor for the phonons and plasmons is shown in panels c) \& e), respectively. The intermediate scattering function is shown in panel d) \& f).  In panel d), the Laplace transform was only performed for energies from $-0.1$~ryd to $0.1$~ryd.}
\end{figure*} 
As before, we also study the case where the bulk of the electrons and protons feature a single temperature, but a fraction of the electrons has a higher temperature, see Fig.~\ref{fig:itcf_theo_7}. The ion modes, as shown in panel c), are hardly affected by the non-equilibrium electron distribution. Thus, if the ITCF is determined by integration over the proton energy scale only, one obtains the temperature of the protons. This is possible here due to the fact that the bulk of the electrons has the same temperature. We again observe the impact of the hot electron population on the dynamic structure in panel e). As seen before, the plasmon position, width and height is altered compared to the equilibrium case. Of particular importance are again the high energy tails of the dynamic structure factor. Thus, the non-equilibrium ITCFs lack symmetry individually and cross the line of unity at different imaginary times $\tau$, see panel f).

Some remarks are in order for the case of a true two-temperature scenario in which the electrons have a temperature and the ions have a different temperature. As already stated in the theory section, the ITCF obtained from the Laplace transform integrated over the electron energy range will then always give the electron temperature. Performing a Laplace transform over just the ion energy range will give a not very useful effective temperature in between the ion and electron temperature because the electronic screening changes the true ion modes. However, if the situation is such that a two-temperature model can be applied in good conscience, then the DPA as introduced in Eq.~(\ref{Lee_no_pi_ei}) is valid. A two-temperature model requires the electron-ion coupling to be weak such that the energy transfer is slow enough to always keep the subsystems in their separate equilibriums. This will be the case on timescales larger than  several hundred femtoseconds and up to several hundred picoseconds. Then, having obtained the total electron structure factor and the temperature of the electrons (we assume the density to be obtained either from the ITCF as well or by other means), Eq.~(\ref{Lee_no_pi_ei}) can be inverted to obtain the pure ion structure ${\cal L}_{ii}$. The latter than allows a Laplace transform for the pure ion ITCF and the ion temperature.
This of course assumes that the ionic energy range of the scattering signal has been resolved sufficiently which is challenging at the moment \cite{descamps_2020,wollenweber_2021}.

\subsection{Time evolution of the scattering signal}
In order to demonstrate the potential of our method to resolve the evolution of the electronic structure via x-ray scattering in a realistic scenario, we use the solutions of 
Monte Carlo simulations
as input for the calculation of the time dependent dynamic structure.

\begin{figure*}
\centering
\includegraphics[width=\textwidth]{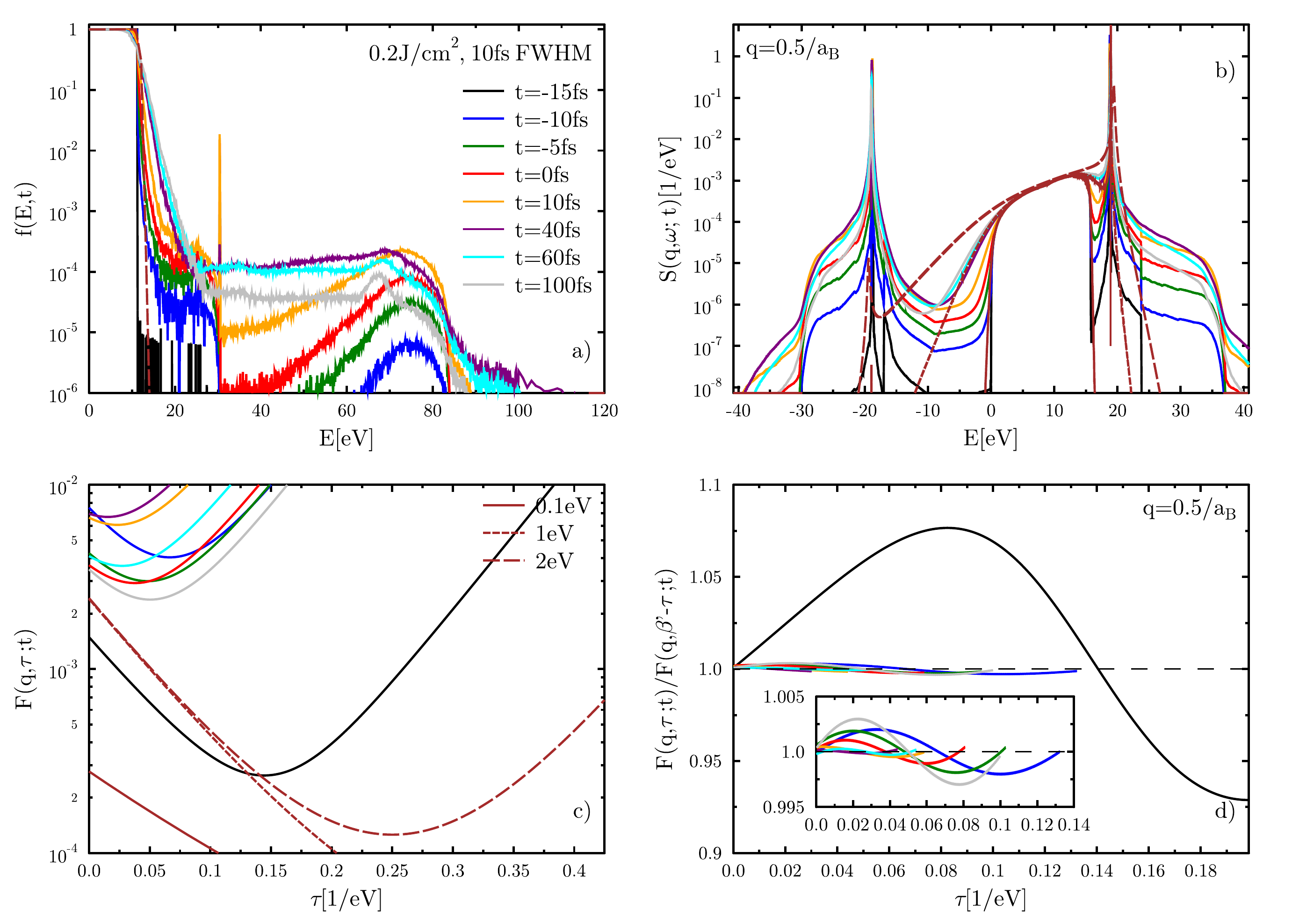}
\caption{\label{fig:itcf_num_1} Time evolution of the non-equilibrium structure factor of the electrons in aluminium under laser irradiation. The electronic density is $n=1.7\times 10^{23}$~cm$^{-3}$. The Wigner distribution functions are shown in panel a). The maximum laser intensity is at $t=0$.  The time dependent dynamic structure factor is shown in panel b). The ITCF is shown in panel c).  In panel d), the ratio between the ITCF of two different imaginary time arguments shows the deviation from the equilibrium symmetry. For comparison, we show equilibrium data for temperatures of $0.1$~eV, $1$~eV, and $2$~eV, respectively (solid, short dashed, and long dashed brown curves).}
\end{figure*} 
Fig.~\ref{fig:itcf_num_1} shows an example of a realistic laser fluence of $0.2$~J/cm$^2$ onto a solid slab of aluminium. The laser FWHM is $10$~fs and the laser wavelength is $135$~\AA~($92$~eV) corresponding to the VUV range of the spectrum. Initially, all electrons are cold and contained in the Fermi function (black) with just some initial few electrons already being scattered into higher states. The smallest time step of $5$~fs in panel a) covers typical changes in the Wigner functions but is smaller than the possible state-of-the-art time resolution for driver, probe laser and detectors at the European XFEL. We observe that the VUV pump laser creates a complicated time dependent electron energy distribution that features several attributes that we have already discussed in detail in the previous sections. 
The dominant photo-absorption for 92-eV-photons in aluminium is by L-shell electrons, which creates the sharp peak at $30$~eV (considered with broadening)~\cite{Medvedev2011a}. A smaller contribution of the photo-absorption by the conduction band electrons promotes photo-electrons to the energies of $\sim 90 - 100$~eV. The bump in the population around $80$~eV is produced predominantly by the Auger-decays of L-shell holes~\cite{Medvedev2011a}. There is a tail in between this bump and the cold bulk being fed by electrons relaxing down from the bump. 

Somewhat surprising, the pump laser also creates a tail of high energy electrons beyond the nominal pump energy due to photoabsorption by the excited electrons (sequential multi-photon absorption), and the scattering of electrons~\cite{Medvedev2011a}. The bump increases in signal strength for as long as the drive laser is on. The high energy tail continues to be more populated even when the laser has ceased to drive further electrons into the bump (purple vs. orange curves). The energy relaxation means that the formerly steep flank of the Fermi distribution around the Fermi energy $\sim 11~eV$ becomes less and less steep for all considered times as the high energy electrons thermalize. It is clear that even after $100$~fs, the electron population has not reached a new equilibrium yet. The bump in the distribution function at $\sim 60-70$ eV is still being fed by the Auger decays of the L-shell. The complete relaxation of XUV-excited electrons in aluminium is thus expected on the scale of a few hundred fs, as supported by experiments~\cite{Bisio2017}. 

Panel b) of Fig.~\ref{fig:itcf_num_1} displays the time resolved dynamic structure factors for a wavenumber in the collective regime. In particular for early times, the dynamic structure contains signatures around $-15$~eV in addition to the collective mode peaks. These are caused by the population of electrons round $20$~eV. Compared to the considered equilibrium examples, the plasmon modes are widened due to the high energy electrons. Temperatures of $1$~eV or $2$~eV have been suggested as final temperatures after equilibration. Overall, if the resolution of the detector is good enough, clear distinction between equilibrium and non-equilibrium is possible. However, the ITCFs show a very symmetric shape [see panels c) and d)]. Apart from the curious case at $t=-15$~fs, the deviation from an equilibrium like symmetry in the ITCF is less than $1\%$ at all times. This is caused, as was discussed with Fig.~\ref{fig:itcf_theo_1} already, by the dominance of the collective modes at this scattering angle which always leads to symmetric ITCFs.

\begin{figure*}
\centering
\includegraphics[width=\textwidth]{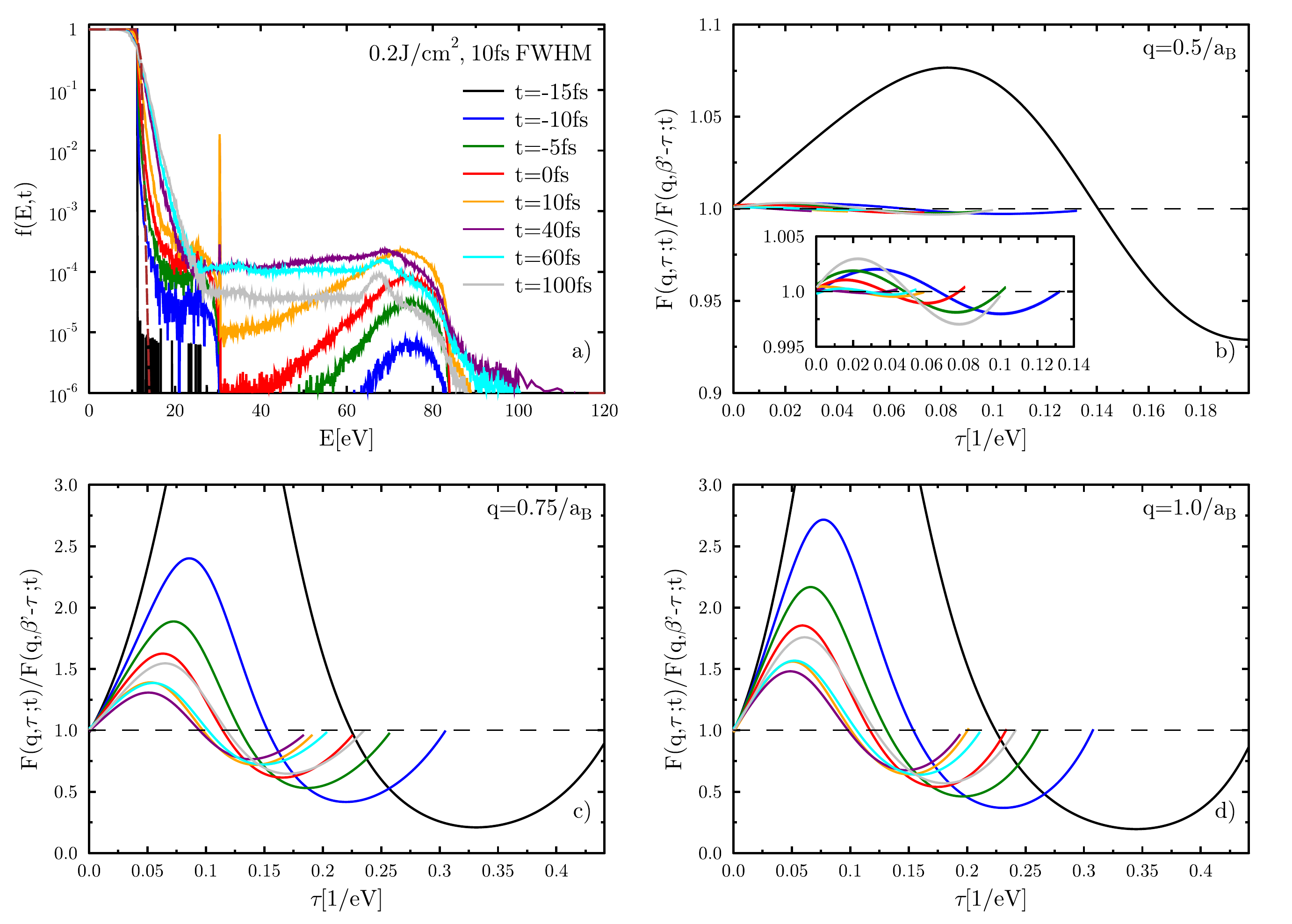}
\caption{\label{fig:itcf_num_2} Time evolution of the non-equilibrium ITCF of the electrons in solid aluminium under laser irradiation. The electronic density is $n=1.7\times 10^{23}$~cm$^{-3}$. The Wigner distribution functions are shown in panel a). In panels b), c) \& d), the ratio between the ITCF of two different imaginary time arguments shows the deviation from the equilibrium symmetry for different wavenumbers.}
\end{figure*} 
Thus, we re-analyze the dynamic structure factor at two different wavenumbers as shown in Fig.~\ref{fig:itcf_num_2}. Panels a) and b) repeat the quantities of Fig.~\ref{fig:itcf_num_1} for comparison. We observe that the small change of the scattering angle to $0.75/$a$_B$ or to $1.0/$a$_B$ instead of $0.5/$a$_B$ is sufficient to see large deviations from the $F(q,\tau)=F(q,\tau-\beta)$ symmetry of equilibrium. Even though the black curve for the smallest time again shows the largest deviation from unity, at later times the ITCF deviation shows the expected behavior: these curves should relax towards unity for larger and larger times. For intermediate times, the connection between different non-equilibrium features of the Wigner distribution and particular deviations from symmetry in the ITCF is somewhat convoluted such that plateaus in the Wigner distribution lead to non-monotonous behavior of the ITCFs for different times. Nevertheless, the power of the ITCF to detect non-equilibrium, especially in the transitional and single particle regime, is obvious.

\begin{figure*}
\centering
\includegraphics[width=\textwidth]{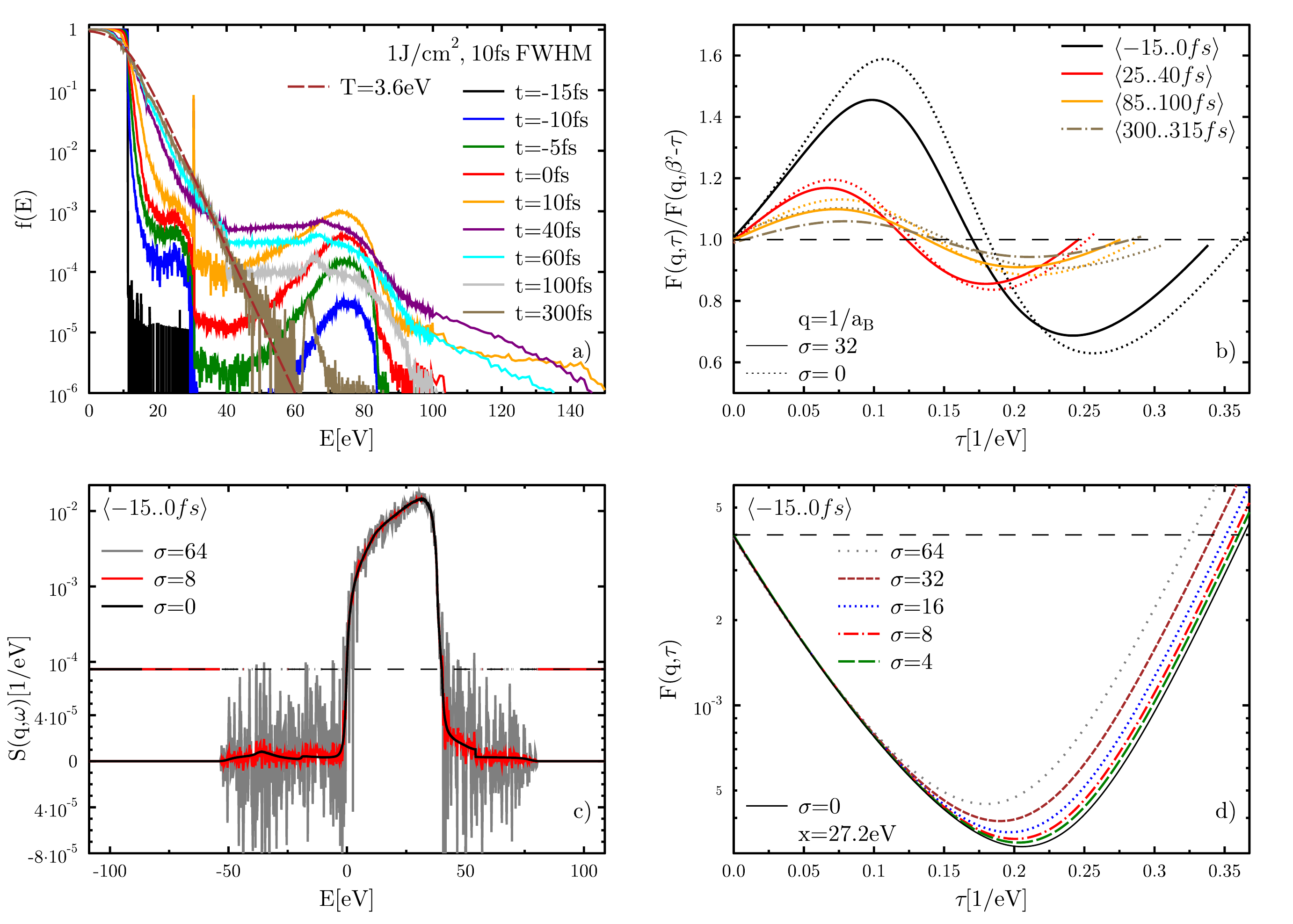}
\caption{\label{fig:itcf_num_3} Time evolution of the non-equilibrium ITCF of the electrons in solid aluminium under laser irradiation of $1$~J/cm$^2$, $10$~fs FWHM, wavelength $\lambda=135$~\AA ($92$~eV) . The electronic density is $n=1.7\times 10^{23}$~cm$^{-3}$, the wavenumber is $q=1/$~a$_B$. The Wigner distribution functions are shown in panel a).  The time dependent averaged ITCF deviations with and without noise are shown in panel b). The scattering signal (dynamic structure factor) and the ITCF  for various noise levels are shown in panels c) \& d) for the case of averaging over the first $15$~fs.  
Note that the y-axis is split into linear and logarithmic scales at $10^{-3}$ in panel c).}
\end{figure*} 

For an example of a realistic signal and analysis of an experiment via the ITCF, we consider the case of a higher laser energy flow onto the target in Fig.~\ref{fig:itcf_num_3}. We observe similar features as before in the electron Wigner distributions. However, due to the higher laser fluence, the cold bulk population becomes severely depleted and the high energy tail becomes stronger. The empty energy band between cold bulk and pump bump is filled with electrons at $t=0$ ($10$~fs earlier than before) and as early as after $t=40$~fs, the Wigner distribution is monotonically decaying for all energies (this process needed more than $100$~fs in the $0.2$~J/cm$^2$ case). 

We have opted to calculate the dynamic structure and the ITCF for a wavenumber at which it will be possible to observe non-equilibrium effects in the ITCF at just this single scattering angle. In order to well represent the actual signal measured in an experiment, we disturb the calculated dynamic structure factors with Gaussian noise of zero mean and the given standard deviation $\sigma$~\cite{sheffield2010plasma,Dornheim_T_follow_up}. We also account for the fact that measurements at modern facilities have a time resolution limited by the XFEL pulse duration~\cite{Berg_2018}; we assume a minimum here of around $15$~fs and average the obtained dynamic structure factors accordingly before adding the noise.  
The energy resolution of the data has been coarse grained to $0.7$\,eV to be of the same order of magnitude as the dispersion of the HAPG spectrometer at HED~\cite{Preston2020}.

The dynamic structure for the first time interval considered is shown in panel c). Several possible noise levels are given. The corresponding ITCFs are shown in panel d) of Fig.~\ref{fig:itcf_num_3}. It can be seen that the different levels of noise are well handled by the Laplace transform. The considered noise does not influence the ability of the ITCF to detect the non-equilibrium since the asymmetric shape of the ITCF remains largely unaltered. 

We have fixed the integration interval for the Laplace transform to $x=\pm 27.2$~eV. This value coincides with a resolution of the expected scattering intensity over three to four orders of magnitude in experiment. If the detection threshold could be expanded to six or seven orders of magnitude, then a detection of the remnants of the high energy electrons between $60$ and $80$~eV at $300$~fs would be possible. In any case, the small deviations from the line at unity still in place in the ITCF at $300$~fs reflects the not perfect Fermi-/Boltzmann-shape of the Wigner function in the small and medium energy range, as can be seen in panel a) of Fig.~\ref{fig:itcf_num_3} when comparing to the best fit at $T=3.6$~eV.

The final analysis of the non-equilibrium state and its time evolution for this particular case of laser irradiation is presented in panel b) of Fig.~\ref{fig:itcf_num_3}. It displays the deviation of the ITCFs from the symmetric equilibrium shape. Solid lines represent the case including noise, dashed lines show results without noise added. There are small deviations between the curves in the two cases, as can be expected. However, the deviation from the symmetric equilibrium case is still well visible when noise is included. Even better, the shape of the deviation from equilibrium remains largely unaltered when noise is considered. The relaxation towards equilibrium is  apparent when comparing these three averaging windows. We can compare to the Wigner distributions of panel a), from which it is clear that after $100$~fs no thermal distribution has been established yet and we need on the order of $300$~fs for a final temperature of around $3.6$~eV to be established. 

We can also observe that electrons of higher energies seem to scatter towards energies in the range $60-80$~eV faster than deficiencies in the population at low energies can be filled. This is the reason that it takes well over $100$~fs to establish a Wigner function that looks even remotely like an equilibrium distribution.




\section{Summary and Outlook\label{sec:summary}}

We have shown the power of the imaginary time correlation function (ITCF) to detect signatures of non-equilibrium in the scattering signal obtained in experiments utilizing VUV, x-ray, or electron probe beams. Further, analyzing the properties of the ITCF as they change over femto-, pico-, or even nano-seconds, a diagnostic of the driving and relaxation of the non-equilibrium state is possible. The entire analysis is model free and is based on the degree of violation of the detailed balance principle that is valid only in equilibrium. 

We concentrated on non-equilibrium features of the Wigner distribution function of the electrons. For these cases, we used model Wigner functions that feature typical artefacts of non-equilibrium, e.g., high energy bumps and tails. We also used solutions of a Boltzmann-MC code to calculate femto-second time resolved scattering spectra and ITCFs for the case of laser heated aluminium. We time-averaged these ITCF spectra and introduced artificial noise to show the capabilities of the method under typical experimental diagnostics conditions. 

Clear indications for non-equilibrium can be seen in all cases for wavenumbers (scattering angles) in the single particle or transitional scattering regime. In such an experimental setup, a single detector is sufficient to trace the signatures of non-equilibrium. In the non-equilibrium collective regime, dominating collective peaks in the dynamic structure factor (plasmons, acoustic modes) will cause the ITCF to be symmetric, even in non-equilibrium, although with a fake 'temperature'. A non-equilibrium signal might still be obtained when analyzing not just the interval $\tau\epsilon[0,\beta']$ but the general symmetry around $\beta'/2$ outside this interval. A safer option is in any case the fielding of two or more spectrally resolving detectors. In this case, if one shows a non-symmetric ITCF, it cannot be equilibrium. If all detectors show a symmetric ITCF but the obtained values of $\beta_i$ differ, it must be non-equilibrium. In such a way, we imagine, different local thermodynamic equilibria of distinct parts of the total scattering volume (as in ICF fusion pellets) will reveal itself.


Thus, we expect this method to become a standard tool in future experimental scattering setups. Since the probe beam, even a seeded x-ray laser beam, has a finite bandwidth and the detector has a limited resolution in energy space, it is absolutely necessary to characterize the source and detector function to at least the same accuracy and precision as the actual scattering signal. In particular for two-temperature relaxation, we expect a qualitative and quantitative improvement in the experimental abilities and characterization of the target, as nowadays often only the ion temperature is measured but not the electron temperature~\cite{ernstorfer,Fletcher_2022}.

\section*{Acknowledgments}
This work was partially supported by the Center for Advanced Systems Understanding (CASUS) which is financed by Germany’s Federal Ministry of Education and Research (BMBF) and by the Saxon state government out of the State budget approved by the Saxon State Parliament. 
NM gratefully acknowledges financial support from the Czech Ministry of Education, Youth and Sports (grants No. LTT17015, LM2018114, and No. EF16\_013/0001552). Computational resources for Boltzmann-MC calculations were supplied by the project "e-Infrastruktura CZ" (e-INFRA LM2018140) provided within the program Projects of Large Research, Development and Innovations Infrastructures.

\bibliography{bibliography}
\end{document}